\documentclass[epjCONF,columns,a4paper]{svjour} %for 2 columns format
\usepackage[pdftex]{graphicx}
\usepackage[varg]{txfonts} % Times fonts
\usepackage[latin1]{inputenc}
\newcommand{\gf}{\ensuremath{G_{\mathrm{F}}}}

\newcommand{\cm}{\ensuremath{\hbox{ cm}}}
\newcommand{\m}{\ensuremath{\hbox{ m}}}

\newcommand{\gev}{\ensuremath{\hbox{ GeV}}}
\newcommand{\tev}{\ensuremath{\hbox{ TeV}}}

\newcommand{\s}{\ensuremath{\hbox{ s}}}

\def\abs#1{\left| #1\right|}
\def\ibid{\hbox{\it ibid.\/}}
\newcommand{\pb}{\hbox{ pb}}
\newcommand{\fb}{\hbox{ fb}}
\newcommand{\smgg}{\ensuremath{\mathrm{SU(3)_c} \otimes \mathrm{SU(2)_L} \otimes \mathrm{U(1)}_Y}}
\newcommand{\cgg}{\ensuremath{\mathrm{SU(3)_c}}}

\newcommand{\ewgg}{\ensuremath{\mathrm{SU(2)_L} \otimes \mathrm{U(1)}_Y}}
\newcommand{\wigg}{\ensuremath{\mathrm{SU(2)_L}}}
\newcommand{\ygg}{\ensuremath{\mathrm{U(1)}_Y}}
\newcommand{\emgg}{\ensuremath{\mathrm{U(1)_{EM}}}}
\newcommand{\lum}{\cm^{-2}\s^{-1}}

\session-title{2011 Hadron Collider Physics Symposium (HCP-2011)}
\begin{document}
\title{Particle Physics in a Season of Change}
\author{Chris Quigg\thanks{\email{quigg@fnal.gov}}\hfill \fbox{\textsf{FERMILAB-CONF-12-035-T}}}
\institute{Fermi National Accelerator Laboratory, P.O. Box 500, Batavia, Illinois 60510 USA}
\abstract{
A digest of my opening remarks at the 2011 Hadron Collider Physics Symposium. 
} %end of abstract
\maketitle
\section{Introduction \label{intro}}
I have chosen my title to reflect the transitions we are living through, in particle physics overall and in hadron collider physics in particular. Data-taking has ended at the Tevatron, with $\sim12\fb^{-1}$ of $\bar{p}p$ interactions delivered to CDF and D\O\ at $\sqrt{s} = 1.96\tev$. The Large Hadron Collider has registered a spectacular first full-year run, with ATLAS and CMS seeing $> 5\fb^{-1}$, LHC$b$ recording $\sim 1\fb^{-1}$, and ALICE logging nearly $5\pb^{-1}$ of $pp$ data at $\sqrt{s} = 7\tev$, plus a healthy dose of Pb-Pb collisions. The transition to a new energy regime and new realms of instantaneous luminosity exceeding $3.5 \times 10^{33}\lum$ has brought the advantage of enhanced physics reach and the challenge of pile-up reaching $\sim 15$ interactions per beam crossing.

I am happy to record that what the experiments have (not) found so far has roused some of my theoretical colleagues from years of complacency and stimulated them to think anew about what the TeV scale might hold. We theorists have had plenty of time to explore many proposals for electroweak symmetry breaking and for new physics that might lie beyond established knowledge. With so many different theoretical inventions in circulation, it is in the nature of things that most will be wrong. Keep in mind that we learn from what experiment tells us is not there, even if it is uncommon to throw a party for ruling something out. Some non-observations may be especially telling: the persistent absence of flavor-changing neutral currents, for example, seems to me more and more an important clue that we have not yet deciphered.

It is natural that the search for the avatar of electroweak symmetry breaking preoccupies participants and spectators alike. But it is essential to conceive the physics opportunities before us in their full richness. I would advocate a three-fold approach: Explore, Search, Measure! The first phase of running at the LHC has brought us to two new lands---in proton-proton and lead-lead collisions---and we may well enter other new lands with each change of energy or increase of sensitivity. I believe that it will prove very rewarding to spend some time simply \emph{exploring} each new landscape, without strong preconceptions, to learn what is there and, perhaps, to encounter interesting surprises. Directed \emph{searches}, for which we have made extensive preparations, are of self-evident interest. Here the challenge will be to broaden the searches over time, so the searches are not too narrowly directed. Our very successful conception of particles and forces is highly idealized. We have a great opportunity to learn just how comprehensive is our network of understanding by making precise \emph{measurements} and probing for weak spots, or finding more sweeping accord between theory and experiment.

Indeed, one of the strengths of our position coming into the new era is that over the past few decades, we have conceived, elaborated, and validated two new laws of nature, quantum chromodynamics and the electroweak theory. These gauge theories derive  interactions among pointlike ($r \lesssim 10^{-18}\m$) quarks and leptons from the \smgg\ symmetry inferred from experimental observations. How the \ewgg\ symmetry is hidden is the most urgent question we face. Let us briefly consider the two components of the standard-model interactions.

\section{Quantum Chromodynamics \label{qcd}}
A defining feature of QCD is that it is an asymptotically free theory in which the running coupling decreases with increasing scale, or decreasing distance~\cite{qcdrl}. Indeed, experiments have established that $\alpha_{\mathrm{s}}(Q)$ decreases from a bit less than $1/3$ in the neighborhood of $Q = 2\gev$ to less than $1/9$ near $Q=200\gev$. It is this property that reconciles the success of the parton model with the non-observation of free quarks, and enables the many successes of perturbative calculations, now reaching to $Q \approx 1\tev$. Notable examples are the agreement between theory and experiment for $p^\pm p \to \hbox{ dijets} + X$  over roughly nine orders of magnitude in cross sections and the resemblance to Rutherford scattering that tests both QCD dynamics and the elementary character of quarks. Thanks to progress in lattice gauge theory, we have also achieved a growing understanding of the nonperturbative regime. Calculations of the light-hadron masses that incorporate the influence of quark-antiquark pairs reproduce the spectrum within a few percent and establish that color confinement---\textit{i.e.,} QCD---explains nearly all of the nucleon mass.

From the point of view of mathematical self-consistency, QCD could be complete up to the Planck scale, but that doesn't prove that it is. While the theory exhibits no structural defects, we still do not have an established solution to the strong \textsf{CP} problem. Even as we regard QCD as a solid basis for calculating backgrounds and signals, we need to prepare for surprises. If we ask how QCD might crack, we might imagine breakdowns of factorization (which would compromise our ability to make reliable perturbative calculations), the observation of free quarks or unconfined color, novel kinds of colored matter, quark compositeness, or a larger symmetry containing QCD. 

It is arguably more likely that we will encounter new phenomena \emph{within} QCD. Examination of (``soft'') multiparticle production might reveal additional components beyond the established diffraction plus short-range-order. The expected high density of few-GeV partons may have several novel consequences, including thermalization (perhaps revealed in high-multiplicity events) and events containing many minijets. Long-range correlations may emerge. I suspect that a few percent of ``minimum-bias'' events may exhibit unusual event structures, with the ``few percent'' increasing with $\sqrt{s}$ and charged-particle multiplicity. Bjorken has suggested that we might be able to recognize collisions involving different configurations of the valence quarks in the proton. To cite a single example, a quark--diquark  body plan for the proton might imply diquark--diquark collisions with characteristics different from the familiar quark--quark collisions. Scanning event displays tailored to the anticipated dynamics of multiple production may be an effective way to explore for hints of new phenomena~\cite{l2c}.
 
ALICE, ATLAS, and CMS have opened a new chapter of heavy-ion physics at the LHC. The clear signs of jet quenching~\cite{Aad:2010bu} and indications of quarkonium ($\Upsilon$) melting~\cite{Chatrchyan:2011pe} are only the first indications of a rich field of study to come. My formula, Explore, Search,  Measure!  applies to heavy-ion collisions as well.

\section{The Electroweak Theory \label{ewth}}
To good approximation, the electroweak theory is specified by a three-generation $\mathrm{V-A}$ structure for the charged-current weak interactions. Flavor-changing neutral currents are suppressed by the Glashow--Iliopoulos--Maiani mechanism, and the (Cabibbo--Kobayashi--Maskawa) quark-mix\-ing matrix describes \textsf{CP} violation~\cite{unanswered}.

Although the underlying \ewgg\ gauge symmetry is broken to \emgg,  LEP experiments have tested the full symmetry by measuring the cross section for the reaction $e^+ e^- \to W^+ W^-$ and confirming the intricate gauge-symmetry cancellation among the $\nu_e$-, $\gamma$-, and $Z^0$-exchange diagrams~\cite{lepewwg}. Assiduous study of quantum corrections to many observables led to the  inference that the top-quark mass must lie in the interval $150\gev \lesssim m_t \lesssim 200\gev$.

Overall, the accord between the electroweak theory and observations is highly impressive, but perhaps the agreement is not perfect~\cite{soni}. Among persistent tensions in $B$ sector, it is worth noting the divergence among inclusive, exclusive, and annihilation determinations of the quark-mixing matrix element  $\abs{V_{ub}}$. An amusing suggestion is that the three different values could be brought into agreement by allowing for a small right-handed $u \leftrightarrow b$ interaction~\cite{Buras:2010pz}.

An unknown \emph{agent provocateur} hides the electroweak symmetry. One of our prime goals is to identify that agent. The possibilities include $\Box$~a force of a new character, based on interactions of an elementary scalar (the Higgs boson of the standard model); $\Box$~a new gauge force, perhaps acting on undiscovered constituents; $\Box$~a residual force that emerges from strong dynamics among electroweak gauge bosons; $\Box$~an echo of extra spacetime dimensions. 

The electroweak theory does not predict the mass of the Higgs boson, but a simple thought experiment identifies a tipping point, or conditional upper bound on $M_H$. The lowest-order diagrams for scattering of $W^+ W^-$, $Z^0 Z^0$, $HH$, and $HZ^0$ satisfy $s$-wave unitarity, provided that 
\begin{equation}
M_H < \left(\frac{8\pi\sqrt{2}}{3\gf}\right)^{1/2} \approx 1\tev.
\label{eq:lqt}
\end{equation}
 If this bound is respected, perturbation theory is reliable (except near particle poles) and the Higgs boson can be observed on the 1-TeV scale.  If not, the weak interactions among $W^\pm$ and $Z^0$ become strong on 1-TeV scale. Either way (and this important conclusion holds beyond the standard electroweak theory), new physics is to be found on the 1-TeV scale. Within the standard model, analyses of quantum corrections favor $114\gev \lesssim M_H \lesssim 143\gev$ at 95\% confidence level~\cite{gfit}. These studies confirm that something like the Higgs boson couples to $W^+W^-$ and $Z^0 Z^0$ as prescribed by the electroweak theory, but they are insensitive to Higgs-boson couplings to fermions. At this moment, \textit{we do not know} that the agent of electroweak symmetry breaking gives mass to fermions. Even if the standard electroweak theory should turn out be correct in every particular, \textit{we do not know} what determines fermion masses and mixings. No calculation within the electroweak theory yields the mass of the electron, or of the top quark, or any relations among quark and lepton masses. I  regard all the quark and lepton masses as evidence for physics beyond the standard model.

Within the next year, our experiments may gain the sensitivity needed to pronounce on the existence or non-existence of the standard-model Higgs boson. That will be a monumental moment in the development of our science, and we must be prepared to explain the motivation and consequences as accurately, completely, and engagingly as we can. That means, first, that you are personally responsible for getting the science right. Some of our colleagues (and too many science writers) persist in repeating the manifestly false statement that the Higgs boson is responsible for all mass, when it is QCD that explains most of the visible mass of the universe in the form of nucleon masses. You are also personally responsible for not propagating made-up history. The people~\cite{ssb} to whom we owe the marvelous insight that spontaneous breaking of a gauge symmetry leads to massive gauge bosons did not set out to understand why matter (in the form of elementary fermions) has mass, and they had nothing to say about the weak interactions. Both the spontaneously broken \ewgg\ theory and the notion that fermion mass could arise from Yukawa couplings of the Higgs scalar to the fermions appear for the first time in Weinberg's 1967 paper~\cite{ws}. If you owe your acquaintance with the history of the electroweak theory to hearsay or to \textit{la presse people}, you can find a reliable narrative in Ref.~\cite{fec}. The real stories are more powerful than  fictional cartoons.

We also have an obligation to explain why the search for the agent of electroweak symmetry breaking justifies the resources invested in the LHC and the experiments, not to mention your own  time and energy. One approach is to ask how different the world would have been, without a Higgs mechanism or a substitute on the real-world electroweak scale~\cite{gedanken}. Think, for simplicity, of one generation of fermions. Without a Higgs vacuum expectation value, the electron and quarks would have no mass. Eliminating the Higgs mechanism does not alter the strong interaction, so QCD would still confine colored objects into hadrons. The gross features of nucleons derived from QCD---such as nucleon masses---would be little changed if the up and down quark masses vanished. If the quarks are massless, the QCD Lagrangian displays an $\mathrm{SU(2)_L\otimes SU(2)_R}$ chiral symmetry that is spontaneously broken near the confinement scale to isospin symmetry by the formation of $\langle\bar{q}q\rangle = \langle\bar{q}_{\mathrm{L}}q_{\mathrm{R}}\rangle + \langle\bar{q}_{\mathrm{R}}q_{\mathrm{L}}\rangle$ condensates. These condensates  couple left-handed and right-handed quarks, giving rise to the effective ``con\-stit\-uent-quark'' masses and  breaking the electroweak symmetry because left-handed and right-handed quarks transform differently under \ewgg. The weak bosons $W$ an $Z$ acquire masses, but they are 2500 times smaller than in the real world: the analogue of the Fermi constant, $G_{\mathrm{F}}$, is enhanced by nearly seven orders of magnitude. Should the proton be stable, or compound nuclei be produced and survive to late times in this alternate universe, the infinitesimal electron mass would compromise the integrity of matter. The Bohr radius of a would-be atom would be macroscopic (if not infinite), so an electron could not be associated with a specific nucleus and valence bonding would have no meaning.  Seeking the agent of electroweak symmetry breaking, we hope to learn why the everyday world is as we find it: why atoms and chemistry and stable structures such as liquids and solids can exist.

Returning to this world, let us suppose that the agent of electroweak symmetry breaking is indeed a light, elementary scalar. Then we will be forced to ask whether $M_H < 1\tev$ makes sense in the framework of quantum field theory. It seems inevitable that if the ultimate theory contains meaningful distant scales---a unification scale or the Planck scale, for example---quantum corrections would tend to pull the Higgs-boson mass up to far higher scales than $1\tev$, unless it is stabilized by a symmetry or a dynamical principle. This is the essence of the hierarchy problem. Supersymmetry could control the quantum corrections by balancing bosonic and fermionic loop contributions. If dynamical symmetry breaking should yield a light Higgs \textit{Doppelg\"{a}nger} (with or without fermion couplings), the composite nature of that stand-in would also damp quantum shifts and bring them under control.

Despite the hints we have for a light ``Higgs boson,'' we have not seen evidence for the new symmetry or dynamics that might serve to make a low mass natural. We can state the mystery in the form of two puzzles:

\noindent
\P~Puzzle \#1: We expect ``New Physics'' on the 1-TeV scale to stabilize Higgs mass and solve the hierarchy problem, but there is no sign of the flavor-changing neutral currents that occur generically in extensions to the standard model. The notion of minimal flavor violation, that the structure of the quark-mixing matrix controls all flavor phenomena, is a name, but not yet an answer. Accordingly, there is great interest in searches for forbidden or suppressed processes that might reveal something about flavor-changing neutral currents.

\noindent
\P~Puzzle \#2: We expect ``New Physics'' on the 1-TeV scale to stabilize Higgs mass and solve the hierarchy problem, but experiment has not established a  pattern of serious quantitative failures of electroweak theory.

To these two puzzles, which have been growing in significance since the LEP era, we may add the observation that no departures from established physics  have turned up in early running of the LHC. Supersymmetry, in particular, is hiding very effectively. Perhaps it is time to ask whether the unreasonable effectiveness of the standard model (to borrow a turn of phrase from Wigner~\cite{wigner}) is itself a deep clue to what might lie beyond the standard model.

One of the ambitious hopes for a more comprehensive takes the form of a unified theory that encompasses the \smgg\ standard model. Simple examples of unified theories foresee a unification of forces in which the (suitably normalized) \cgg, \wigg, and \ygg\ coupling constants evolve toward a common value at some high unification scale. Calculations in perturbation theory, applied to the measured low-energy values of the couplings, show that coupling-constant unification is more promising in supersymmetric $\mathrm{SU(5)}$ than in the original $\mathrm{SU(5)}$ theory, provided that the change in evolution due to a full spectrum of superpartners occurs near $1\tev$. 

\begin{figure}
  \centerline{\includegraphics[width=0.95\columnwidth]{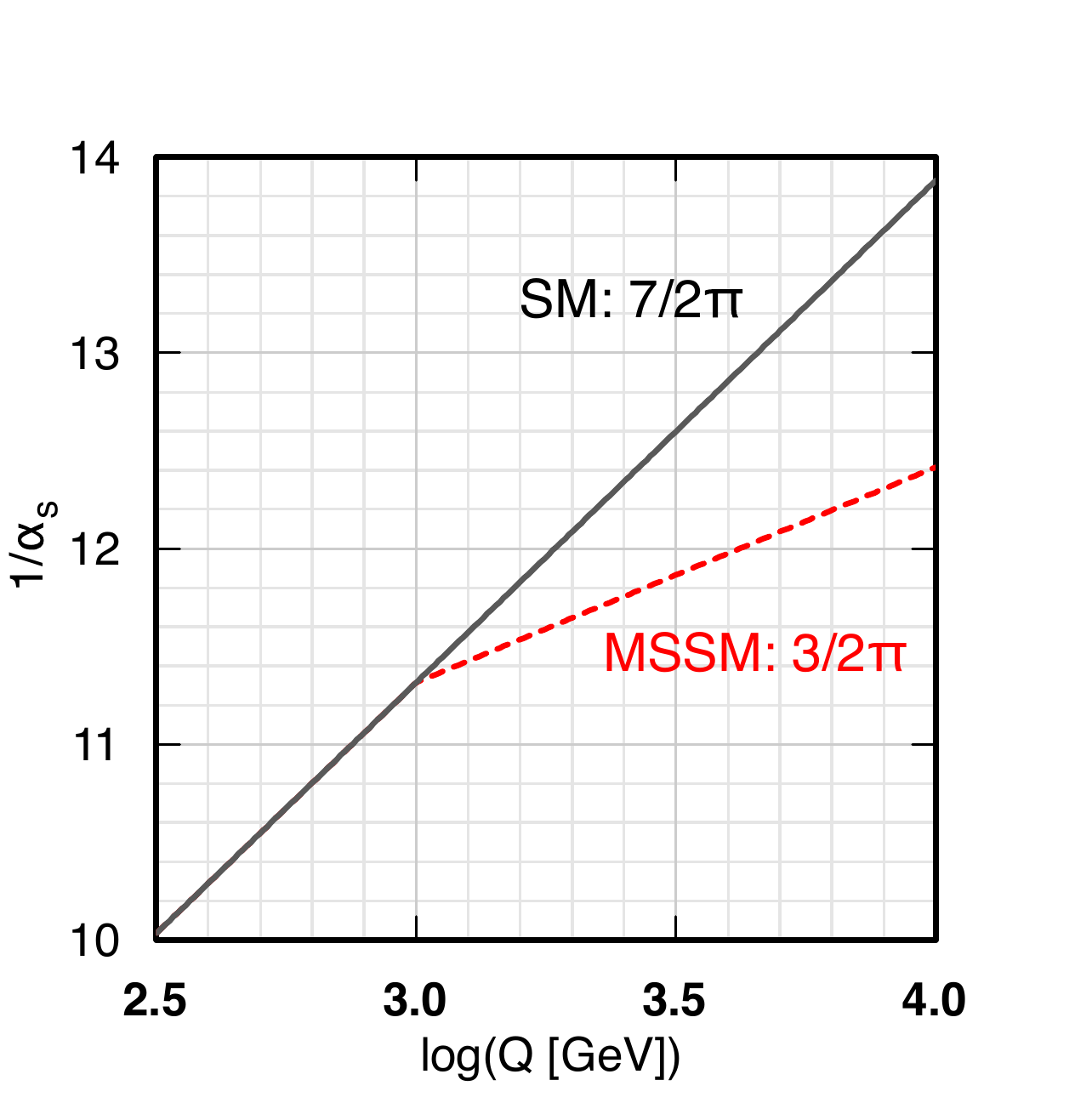}} 
\caption{Evolution of the strong coupling constant (displayed as $1/\alpha_{\mathrm{s}}$) in the standard and supersymmetric versions of the $\mathrm{SU(5)}$ unified theory, for superpartners that become active at $Q = 1\tev$.}
\label{fig:alphasbreak}       
\end{figure}

It is interesting to ask whether LHC experiments could test this hypothesis by measuring the strong coupling constant $\alpha_{\mathrm{s}}$ or the weak mixing parameter $\sin^2\theta_{\mathrm{W}}$ as functions of scale. I sketch in Figure~\ref{fig:alphasbreak} the evolution of $1/\alpha_{\mathrm{s}}$, in leading logarithmic approximation, with and without a superpartner threshold at $Q = 1\tev$.
The slope changes significantly, from $7/2\pi$ to $3/2\pi$, at what I have taken here as a sharp threshold. Seeing, or not seeing, such a change would be powerful evidence for or against the existence of a new set of colored particles that would complement ongoing searches for specific new-particle signatures.

Considerable work will be required to determine promising classes of measurements. I suspect that the study of $Z^0 + \hbox{jets}$ will be fruitful. A continuing dialogue between theory and experiment will be needed to isolate $\alpha_{\mathrm{s}}(Q)$ measured at a high scale.

\section{\`{A} suivre \ldots \label{coming}}
\begin{figure}
  \centerline{\includegraphics[width=0.95\columnwidth]{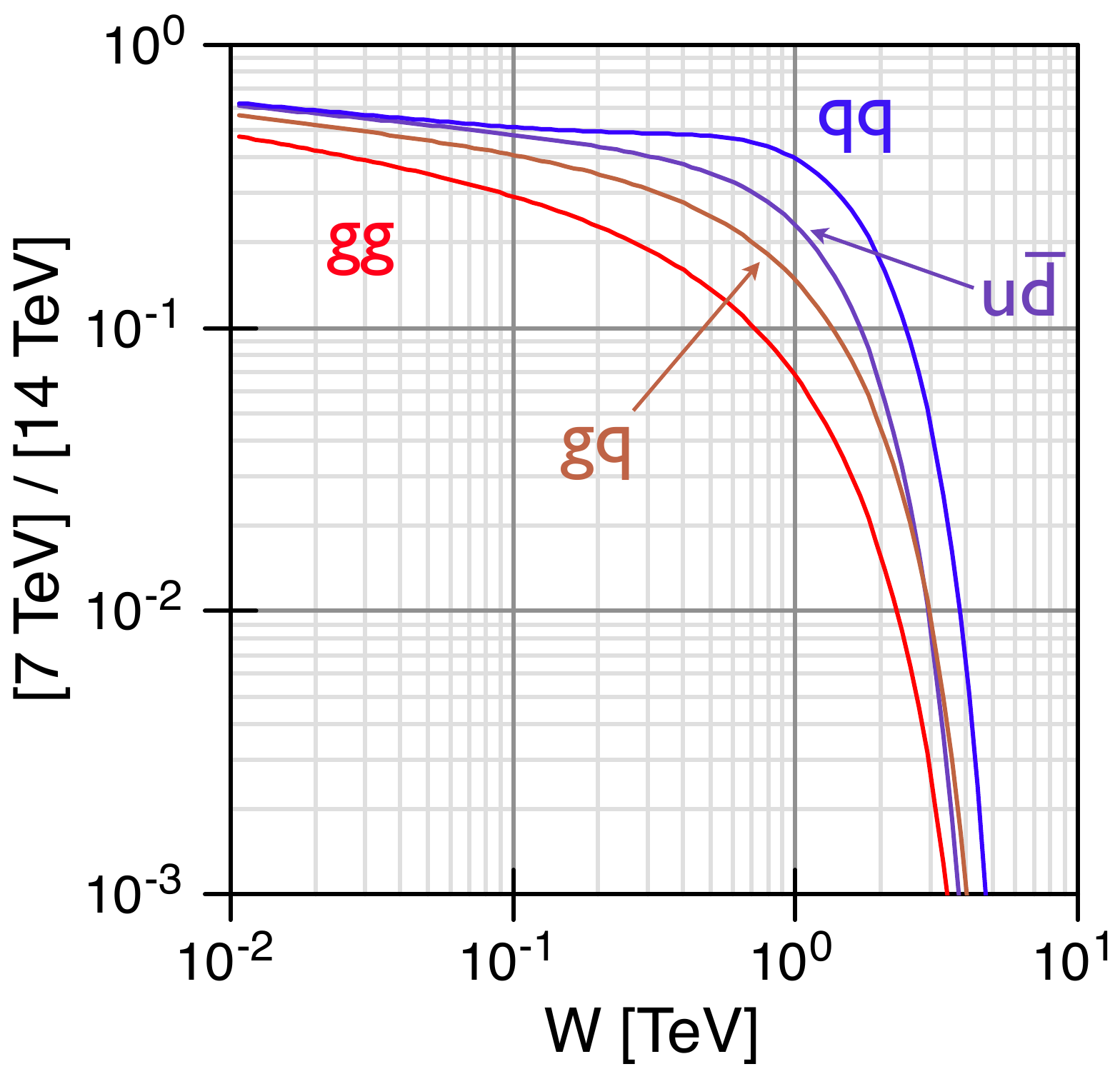}} 
\caption{Ratios of parton luminosities in $pp$ collisions at $\sqrt{s}=7\tev$ and $14\tev$. At $W=1\tev$, the curves from top to bottom refer to $qq$, $u\bar{d}$, $gq$, and $gg$, where $q$ is a light ($u\hbox{ or }d$) quark. The CTEQ6L1 parton distributions are used. (From Ref.~\cite{partonlum}.)}
\label{fig:partonlum}       
\end{figure}

The last events have been recorded at the Tevatron collider, but the interpretation of data, now enriched by conversation with LHC experiments, continues. I look forward to a rich year of final results from CDF and D\O, ranging from searches for the standard-model Higgs boson to legacy measurements of the $W$-boson mass. The LHC experiments have made wonderful beginnings on the three fronts---Explore, Search,  Measure!---but we have miles to go. The impressive luminosities delivered to ALICE, ATLAS, CMS, and LHC$b$ prefigure the much larger data samples to come, and the LHC is operating at only half its design energy. I show in Figure~\ref{fig:partonlum} the ratios of selected parton luminosities at $\sqrt{s} = 7\tev$ to those at $\sqrt{s}=14\tev$, as a function of the parton-parton subenergy $W$. Even in the 100-GeV range, the advantage of $14\tev$ over $7\tev$ is noticeable. For $W \gtrsim 1\tev$, the advantage of higher-energy running becomes decisive.

Here are some of the questions on my mind:
\begin{enumerate}
 \item What is the agent of electroweak symmetry breaking? Is there a Higgs boson? Might there be several? 
\item  Is the Higgs boson elementary or composite? How does it interact with itself? What shapes the Higgs potential---or, more generally, triggers electroweak symmetry breaking?
\item Does the Higgs boson give mass to fermions, or only to the weak bosons? What sets the masses and mixings of the quarks and leptons? (How) is fermion mass related to the electroweak scale?
 \item Will new flavor symmetries give insights into fermion masses and mixings?
 \item What stabilizes the Higgs-boson mass below 1 TeV?
 \item  Do the different charged-current behaviors of left-handed and right-handed fermions reflect a fundamental asymmetry in nature's laws?
\item  What will be the next symmetry we recognize? Are there additional heavy gauge bosons? Is nature supersymmetric? Is the electroweak theory contained in a grander unified theory?
\item  Are all flavor-changing interactions governed by the standard-model Yukawa couplings? Does ``minimal flavor violation'' hold? If so, why? At what scale?
\item  Are there additional sequential quark and lepton generations? Or new exotic (vector-like) fermions?
\item What resolves the strong \textsf{CP} problem?
 \item  What are the dark matters? Might dark matter have a flavor structure?
\item  Is electroweak symmetry breaking an emergent phenomenon connected with strong dynamics? How would that alter our conception of unified theories of the strong, weak, and electromagnetic interactions?
\item  Is electroweak symmetry breaking related to gravity through extra spacetime dimensions?
\item What resolves the vacuum energy problem?
\item (When we understand the origin of electroweak symmetry breaking,) what lessons will electroweak symmetry breaking hold for unified theories? \ldots for inflation? \ldots for dark energy?
 \item  Will experiments reveal unexpected  phenomena in strong interactions?
\item  What explains the baryon asymmetry of the universe? Are there new \textsf{CP}-violating phases in charged-current interactions?
\item Are there new flavor-preserving phases? What would observation, or more stringent limits, on electric-dipole moments imply for theories beyond the standard model?
\item (How) are quark-flavor dynamics and lepton-flavor dynamics related (beyond the gauge interactions)? 
\item At what scale are the neutrino masses set? Is the neutrino its own antiparticle? \\
\phantom{\qquad}{\ldots and finally \ldots}
 \item \emph{How are we prisoners of conventional thinking?}
\end{enumerate}
We should have much to digest---and to celebrate---at HCP 2012 in Kyoto. In anticipation of another intense year, I leave you with the words of Stewart Brand~\cite{WholeEarth}:
\begin{quote}
Stay hungry. Stay foolish.
\end{quote}

\section*{Acknowledgements}
Fermilab is operated by Fermi Research Alliance, LLC  under Contract
No.~DE-AC02-07CH11359 with the United States Department of Energy.  I thank Gregorio Bernardi and the Organizing Committee for the invitation to speak, and thank all participants for a lively and stimulating symposium.

%\begin{figure}
%  \centerline{\includegraphics[width=0.95\columnwidth]{Plum37r.pdf}} 
%\caption{Ratios of parton luminosities in $pp$ collisions at $\sqrt{s}=7\tev$ and $14\tev$. At $W=1\tev$, the curves from top to bottom refer to $qq$, $u\bar{d}$, $gq$, and $gg$, where $q$ is a light ($u\hbox{ or }d$) quark.}
%\label{fig:partonlum}       % Give a unique label
%\end{figure}
%\begin{table}
%\caption{Please write your table caption here.}
%\begin{tabular}{lll}
%\hline\noalign{\smallskip}
%\end{tabular}
%\end{table}
%
%\frenchspacing

\end{document}